\documentclass{iopart}

\usepackage{lscape}
\usepackage{setspace}
\usepackage{multicol}
\usepackage{enumerate}
\usepackage{subfig}
\usepackage{paralist}
\usepackage{graphicx}
\usepackage{calc,xcolor}
\usepackage{mathptmx,bm}
\usepackage[breaklinks=true]{hyperref}
\usepackage{iopams}

\begin{document}

\topical{Disorder and electronic transport in graphene}
%\title[Disorder in graphene]{Disorder and electronic transport in
%  graphene}

\author{E R Mucciolo$^1$ and C H Lewenkopf$^2$}

\address{$^1$ Department of Physics, University of Central Florida,
  Orlando, FL 32816-2385, USA}

\address{$^{\, 2}$ Instituto de F\a'{i}sica, Universidade Federal
  Fluminense, 24210-346 Niter\a'{o}i, RJ, Brazil}

\ead{\mailto{mucciolo@physics.ucf.edu} and \mailto{caio@if.uff.br}}

%%%%%%%%%%%%%%%%%%%%%%%%%%%%%%%%%%%%%%%%%%%%%%%%%%%%%%%%%%%%%%%%%%%
\begin{abstract}
In this review, we provide an account of the recent progress in
understanding electronic transport in disordered graphene
systems. Starting from a theoretical description that emphasizes the
role played by band structure properties and lattice symmetries, we
describe the nature of disorder in these systems and its relation to
transport properties. While the focus is primarily on theoretical and
conceptual aspects, connections to experiments are also
included. Issues such as short versus long-range disorder,
localization (strong and weak), the carrier density dependence of the
conductivity, and conductance fluctuations are considered and some
open problems are pointed out.
\end{abstract}
%%%%%%%%%%%%%%%%%%%%%%%%%%%%%%%%%%%%%%%%%%%%%%%%%%%%%%%%%%%%%%%%%%%

%\pacs{72.80.Vp,81.05.ue,72.10.Fk,73.22.Pr}

\submitto{J. Phys.: Condens. Matter {\bf 22}, 273201 (2010)}

%\maketitle

%%%%%%%%%%%%%%%%%%%%%%%%%%%%%%%%%%%%%%%%%%%%%%%%%%%%%%%%%%%%%%%%%%%
%% Introduction

% Introduction
%

\section{Introduction}
\label{sec:introduction}
%%%%%%%%%%%%%%%%%%%%%%%%%%%%%%%%%%%%%%%%%%%%%%%%%%%%%%%%%%%%%%%%%%%

In just the few years since its first synthesis
\cite{novoselov05,berger04,zhang05}, graphene has become the center of
attention of a substantial body of scientists in the fields of
physics, materials science, chemistry, and electrical engineering. It
is a fascinating material, made of a single atomic layer of carbon,
with unique electrical, thermal, and mechanical properties. As such,
it is likely to have a large number of practical applications. Even
though the main features of its electronic band structure were
described more than sixty years ago \cite{wallace47}, only in recent
years we started to appreciate (and understand) the wealth of
phenomena in which charge carriers in graphene can participate. There
are already a couple comprehensive reviews in the literature covering
the fundamentals of graphene
\cite{castroneto09,peres10,dassarma10,abergel10} and the reader is
strongly encouraged to consult them for acquiring a background on the
subject. Here we focus instead on how disorder in its various forms
affects the electronic transport properties of graphene at low
temperatures. This topic has become a vast subfield in its own
right. Thus, rather than provide an exhaustive review, in the spirit
of a topical review we try to cover just enough material to bring the
reader up-to-date with key concepts and crucial results. As a
consequence, to narrow the scope, we leave out topics such as ac
transport, quantum Hall effect, and spin transport, and the effect of
electron-electron interactions. Due to the limited space, we also do
not discuss electronic transport in graphene nanoribbons.

The paper is organized as follows. In section \ref{sec:symmetry} we
describe the role of symmetries and symmetry breaking by disorder in
graphene. The nature and differences among the various sources of
disorder in graphene are discussed in section \ref{sec:disorder}. In
section \ref{sec:neutral} we look at transport properties at the
neutrality or Dirac point, while in section \ref{sec:mesoscopics} we
discuss weak localization and conductance fluctuations caused by
quantum interference. The case of doped (non neutral) graphene is
considered in section \ref{sec:doped}. Finally, in section
\ref{sec:conclusions} we draw some conclusions and point to a number
of open questions. Although our emphasis is on theoretical aspects,
connections to experimental results are made throughout the paper.

%%%%%%%%%%%%%%%%%%%%%%%%%%%%%%%%%%%%%%%%%%%%%%%%%%%%%%%%%%%%%%%%%%%

%%%%%%%%%%%%%%%%%%%%%%%%%%%%%%%%%%%%%%%%%%%%%%%%%%%%%%%%%%%%%%%%%%%
%% Symmetry

% The role of symmetry
%

\section{Pristine graphene and the role of symmetry}
\label{sec:symmetry}
%%%%%%%%%%%%%%%%%%%%%%%%%%%%%%%%%%%%%%%%%%%%%%%%%%%%%%%%%%%%%%%%%%%

\begin{figure}[ht]
\centering
\includegraphics[width=9cm]{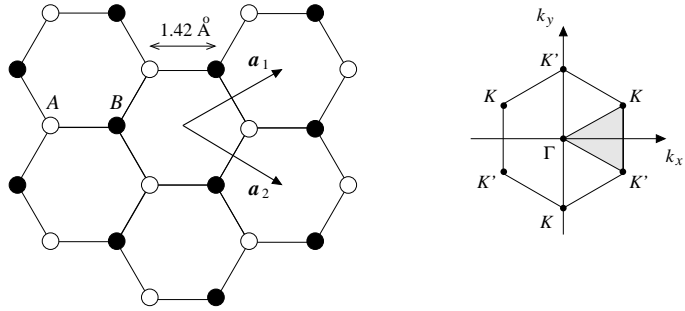}
\caption{Honeycomb lattice and the corresponding Brillouin
  zone. Lattice vectors are denoted by $\bi{a}_1$ and
  $\bi{a}_2$. Empty and solid circles represent different
  sublattices.}
\label{fig:lattice}
\end{figure}

To understand how disorder affects electronic transport properties in
graphene, it is important to consider the ideal case of a pristine
crystal \cite{saito98}. Single-layer graphene is formed by carbon
atoms disposed in a two-dimensional non-Bravais honeycomb lattice
(namely, two intersecting hexagonal sublattices - see figure
\ref{fig:lattice}). The distance between nearest-neighbor sites is
approximately 1.42 \AA, while the lattice constant is $a=2.46$
\AA. The geometry and flatness of the lattice prohibit any overlap
between the $p_z$ orbital of a given atom and the s, p$_x$, and p$_y$
orbitals of its neighbors. The orbitals s, p$_x$, and p$_y$ hybridize
to create the sp$^2$ bonds that hold the atoms together, as well as a
high-energy $\sigma$ band. The so-called $\pi$ band is created by the
overlap of p$_z$ orbitals and can be treated independently from other
bands. In graphene, the $\pi$ band is responsible for most of the
electron conduction. Thus, at low energies or doping, a single-band
tight-binding model including only nearest-neighbor hoping between
adjacent sites on distinct hexagonal sublattices (which are named
${\rm A}$ and ${\rm B}$ hereafter) provides a good approximation for
study of electronic properties in graphene. The hopping amplitude in
this case is $t\approx 2.7$ eV \cite{castroneto09}.

%%%%%%%%%%%%%%%%%%%%%%%%%%%%%%%%%
%%%%%%%%%%%%%%%%%%%%%%%%%%%%%%%%%

When switching to a momentum representation, in addition to the spin
degree of freedom, one has to introduce an isospin structure in the
electron wavefunction to account for the two sublattices. The
resulting Hamiltonian can be diagonalized to yield the energy band
\cite{wallace47}
\begin{equation}
\label{eq:piband}
E_\pm(\bi{k}) = \pm t\, \sqrt{1 + 4\, f(\bi{k})},
\end{equation}
where
\begin{equation}
\label{eq:f}
f(\bi{k}) = \cos^2 ( k_x a/2) + \cos(\sqrt{3}k_x a/2) \cos (k_y a/2).
\end{equation}
Neutral graphene has its Fermi energy at $E_{\rm F}=0$, cutting the
$\pi$ band exactly at the six corners of the hexagonal Brillouin
zone. As shown in figure \ref{fig:lattice}, only two sets of
nonequivalent corner points exist and are denoted by ${\rm K}$ and
${\rm K}^\prime$, with ${\rm K}$ and ${\rm K}^\prime$ points
alternating along the hexagon. Near any of these points, say ${\rm
  K}$, the two (positive and negative) branches of the $\pi$ band in
(\ref{eq:piband}) can be approximated as
\begin{equation}
\label{eq:linearband}
E_\pm(\bi{k}) \approx \pm \hbar v_{\rm F}\, |\bi{k} - \bi{K}|,
\end{equation}
where $v_{\rm F} = \sqrt{3}at/2\hbar \approx 10^6$ m/s is the Fermi
velocity. This is a remarkable result: near $E=0$ (the so-called Dirac
point), carriers in graphene have a linear dispersion relation, much
like massless relativistic particles. In addition, we can introduce
another isospin index to the wavefunction in order to differentiate
contributions from the ${\rm K}$ and ${\rm K}^\prime$ points (also
known as the two valleys), yielding the four-dimensional spinor
\begin{equation}
\Psi = \left( \begin{array}{c} \psi_{\rm A}^{\rm K} \\ \psi_{\rm
    B}^{\rm K} \\ \psi_{\rm B}^{\rm K^\prime} \\ \psi_{\rm A}^{\rm
    K^\prime} \end{array} \right)
\end{equation}
(real spin indices are omitted). Close to the Dirac point, the
effective Hamiltonian acting on this four-dimensional spinor reads
\begin{eqnarray}
\label{eq:Hamiltonian}
H & = & v_{\rm F} \left({\bm p} \cdot \bsigma \right)\, \tau_z
\nonumber \\ & = & \hbar v_{\rm F} \left( \begin{array}{cccc} 0 & k_x
  - i k_y & 0 & 0 \\ k_x + i k_y & 0 & 0 & 0 \\ 0 & 0 & 0 & -k_x + i
  k_y \\ 0 & 0 & -k_x - i k_y & 0 \end{array} \right),
\end{eqnarray}
where the linear momentum $\bi{p}=\hbar(k_x,k_y)$ is defined with
respect to the corner points of the Brillouin zone and the Pauli
matrices $\bsigma=(\sigma_x,\sigma_y)$ and $\tau_z$ act on the ${\rm
  AB}$ and ${\rm KK}^\prime$ subspaces, respectively. The eigenstates
of equation (\ref{eq:Hamiltonian}),
\begin{equation}
\label{eq:planewaves}
\!\!\!\!\!\!\!\!\!\!\!\!\!\!\!\!\!\!\!\!\!\!\!\!\!\!\!\!\!\!\!\!\!\!\!
\!\!\!\!\!\!\!
\left( \begin{array}{c} \psi^{\rm K}_{\rm A} \\ \psi^{\rm K}_{\rm
    B} \end{array} \right) = \frac{1}{\sqrt{2}}e^{i {\bm k}\cdot {\bm
    r} } \left(\begin{array}{c} e^{-i\theta_{\bm k}/2} \\ \pm
  e^{i\theta_{\bm k}/2}
\end{array}\right) \quad {\rm and} \quad
\left( \begin{array}{c} \psi^{\rm K^\prime}_{\rm B} \\ \psi^{\rm
    K^\prime}_{\rm A} \end{array} \right) = \frac{1}{\sqrt{2}}e^{i
  {\bm k}\cdot {\bm r}} \left(\begin{array}{c} \pm e^{-i\theta_{\bm
      k}/2} \\ e^{i\theta_{\bm k}/2}
\end{array} \right),
\end{equation}
provide a basis for solving transport problems in the presence of
disorder. Here, $\theta_{\bm k} = \arctan(k_y/k_x)$ and the $+/-$
signs correspond to the conduction/valence bands, respectively,

One can use the eigenstates in equations (\ref{eq:planewaves}) to
calculate the conductance in pristine graphene. The results are
strongly dependent on geometry and edge orientation
\cite{katsnelson06,tworzydlo06}. For later comparison, let us briefly
comment on these results. The standard ballistic model considers a
graphene ribbon of width $W$ and length $L$ connected to heavily doped
graphene leads. In this geometry, the transverse momentum is quantized
and free propagating modes exist at the contacts. This allows one to
calculate the $S$-matrix and obtain the conductance $G$ using the
Landauer formula. The conductivity, defined as $\sigma=(L/W)G$, has a
minimum at the Dirac point described by
\begin{equation}
\label{eq:pristineconductivity}
\sigma \approx \frac{4e^2}{\hbar} \left( \frac{L}{W} \right) \left[
  g_0 + \sum_{n=1}^{N\gg 1} \frac{1}{\cosh^2(\pi nL/W)} \right],
\end{equation}
where $g_0 = 1/2$ for metallic ribbons, and $g_0=0$ for semiconducting
ones \cite{tworzydlo06}. For $W \approx L$, $\sigma$ is dominated by
$g_0$. In general, $\sigma$, as given by equation
(\ref{eq:pristineconductivity}), depends on the aspect ratio
$W/L$. Hence, calling it conductivity is somewhat misleading, since
the conductivity is usually expected to be a sample specific quantity
and independent of geometry. Yet, for $W/L\gg 1$, $\sigma$ is
dominated by evanescent modes and one finds
\begin{equation}
\sigma \approx \frac{4 e^2}{\pi h} \;.
\end{equation}
This value is often (mis)quoted as the universal ballistic
conductivity minimum, even though its derivation is only applicable to
samples with a narrow aspect-ratio. The Landauer formula can also be
used to calculate $\sigma$ as a function of the carrier concentration
$n$ since the latter is related to the Fermi energy: $n=E_{\rm
  F}|E_{\rm F}|/\pi(\hbar v_{\rm F})^2$. For sufficiently large values
of $|n|$, $\sigma$ reads \cite{muller09}
\begin{equation}
\label{eq:ballistic}
\sigma(n) = \frac{e^2}{h} L \sqrt{\pi |n|}.
\end{equation}

Going back to equation (\ref{eq:Hamiltonian}), we notice that the
effective Hamiltonian of pristine graphene is invariant under a large
number of symmetry operations in the isospin spaces besides the usual
spatial translation and rotation invariances
\cite{ostrovsky07}. First, the Hamiltonian commutes with the operators
\begin{equation}
\Lambda_x = \sigma_z\tau_x, \quad \Lambda_y = \sigma_z\tau_y, \quad
\Lambda_z = \sigma_0\tau_z,
\end{equation}
which, together with $\Lambda_0 = \sigma_0\tau_0$, form a group (here,
the subscript $0$ denotes the identity operator). Second, the
Hamiltonian is also time-reversal and chiral symmetric; the
corresponding operators are denoted by $T_0 = \sigma_x \tau_x$ and
$C_0 = \sigma_z \tau_0$, respectively. Notice that time-reversal is
also accompanied by an inversion in the sign of the linear momentum as
well as a transposition of the Hamiltonian matrix (it basically takes
${\rm K}$ to ${\rm K}^\prime$ and vice-verse). A total of fifteen
distinct symmetry operations can be constructed by combining
$\Lambda_{0,x,y,z}$, $T_0$, and $C_0$. For instance, $\left(T_0
\Lambda_x \right)^{-1} H \left(T_0 \Lambda_x \right) = H$ and
$\left(C_0 \Lambda_x \right)^{-1} H \left(C_0 \Lambda_x \right) = -H$
(the latter exchanges $\pi$ band branches but preserves the energy
spectrum).

Disorder will not only break translation and rotation symmetries but
also affect these other invariances. Thus, to describe static disorder
in general, we add to the Hamiltonian in (\ref{eq:Hamiltonian}) a term
of the form
\begin{equation}
\label{eq:Hdis}
H_{\rm dis} = \sum_{ij} V_{ij} \sigma_i\, \tau_j.
\end{equation}
Depending on which matrix elements $\{V_{ij}\}$ are nonzero, certain
pseudospin symmetries are broken while others are preserved. For
instance, a generic nonmagnetic disorder may break all symmetries but
time-reversal. In fact, the study of the effect of disorder on the
transport properties of two-dimensional massless fermions described by
Hamiltonians of the form (\ref{eq:Hamiltonian}) actually precedes
graphene, and some important facts have been known for quite some time
within the contexts of the integer Quantum Hall effect and d-wave
superconductivity. Let us discuss a few important cases.

Chiral symmetry is associated to the block off-diagonal form of $H$ in
equation (\ref{eq:Hamiltonian}). Therefore, any disorder that
introduces diagonal matrix elements or even a finite chemical
potential (moving the Fermi energy away from the Dirac point) will
break this symmetry and, in most cases, strongly affect the electrical
conductivity. However, whenever $V_{0j} = V_{zj} = 0$, quantum
corrections to the conductivity cancel each other to all orders in the
disorder strength and there is no localization at the Dirac point
\cite{gade93,guruswamy00,mudry03}. The conductivity in this case takes
a finite value which may depend on the disorder strength and whether
the remaining symmetries of the Hamiltonian are broken or not
\cite{ostrovsky06,ostrovsky07}. Bond disorder due to lattice
distortions such as ripples, random magnetic fields, and dislocations
and other lattice defects represented by non-Abelian gauge fields
\cite{morpurgo06} fall into such a class of disorder.

When chiral symmetry is not present, two important situations
arise. First, if the disorder is sufficiently smooth, varying
significantly only over length scales larger than the lattice
constant, there is very little mixing between the two valleys; the
${\rm K}$ and ${\rm K}^\prime$ points remain essentially decoupled and
$V_{ix} = V_{iy} = 0$. For some time in the literature, the fate of
massless Dirac fermions in the presence of long-range scalar disorder
was unclear \cite{fradkin86,ludwig94}.  In fact, some issues appeared
even in the clean limit. For instance, a calculation of the Kubo
conductivity of non-disordered Dirac fermions in the so-called
$\pi$-flux phase model disagreed with the Einstein conductivity
calculated in the thermodynamic limit \cite{ludwig94}. Only recently
an explanation for this discrepancy was provided \cite{ryu07a}, one
reason being that since the neutral point is actually a critical
point, the value of the clean conductivity depends on the order of
limits. This indicates that one should also be careful when
considering claims of universality in the disordered case. However, as
we will discuss below, it became widely accepted recently that
localization due to quantum interference is also absent in this
case. Second, when disorder has a strong short-range component and
intervalley mixing is significant, the opposite occurs and quantum
corrections tend to normalize the conductivity to zero, turning
graphene into an Anderson insulator
\cite{khveshchenko06,aleiner06,altland06,cresti08}. In this case, as
far as transport is concerned, charge carriers in graphene behave much
like those in an ordinary two-dimensional electron gas.

Finally, it is worth remarking that far away from the Dirac point,
graphene is no longer described by the Hamiltonian
(\ref{eq:Hamiltonian}) since the dispersion relation ceases to be
conic. For instance, trigonal warping changes the role of quantum
corrections to the conductivity and enhances localization
\cite{mccann06}. We return to this issue in section
\ref{sec:mesoscopics}, when we address mesoscopic effects in graphene.

%%%%%%%%%%%%%%%%%%%%%%%%%%%%%%%%%%%%%%%%%%%%%%%%%%%%%%%%%%%%%%%%%%%

%%%%%%%%%%%%%%%%%%%%%%%%%%%%%%%%%%%%%%%%%%%%%%%%%%%%%%%%%%%%%%%%%%%
%% Disorder

% The nature of disorder in graphene
%

\section{Nature of disorder in graphene}
\label{sec:disorder}
%%%%%%%%%%%%%%%%%%%%%%%%%%%%%%%%%%%%%%%%%%%%%%%%%%%%%%%%%%%%%%%%%%%

It is commonly accepted that graphene has very few lattice
defects. Therefore, intrinsic disorder tends to be weak even in
exfoliated samples. However, extrinsic disorder is invariably present
and is basically dictated by the synthesis method and by way the
graphene sheet is supported. Chemical contamination in the form of
adsorbates can occur during lithographic processing of field-effect
devices. The focused electron beam used in the lithographic process,
and imaging with SEM also introduce disorder, although the its exact
effect on the graphene lattice is still under debate
\cite{teweldebrhan09,krasheninnikov10}. In addition, when graphene is
laid on a substrate, lattice distortions (the so-called ripples) can
appear due to a tendency in graphene to conform to the roughness in
the substrate surface \cite{ishigami07}. It has been recently
suggested that wrinkles and other singularities can also be formed,
causing lattice distortions \cite{pereira10}. Finally, most insulating
substrates used in graphene devices are oxides which are rather prone
to charge traps. These traps can be located either in the bulk of the
substrate (but not too far away) or at the interface between the
substrate and the graphene sheet and are a common source of disorder.

For clean, suspended graphene systems, extrinsic disorder can be
substantially reduced \cite{bolotin08,du08}. Nevertheless, some amount
of disorder is still present, as shear and strain created by contacts
and scaffolds typically induce corrugation in the graphene sheet
\cite{bao09} which can create electron scattering
\cite{fasolino07,kim08,khveshchenko08,cortijo09,prada09,pereira09a,ribeiro09}. Ripples
induced by thermal motion are rather unlikely at low temperatures, but
can be an important source of scattering at high temperatures in the
form of flexural phonon modes.

It was recognized early that screening in graphene is rather poor due
to the low density of states near the neutrality point. Thus, at low
doping, charge traps in the substrate as well as Coulomb impurities
(e.g. charged adsorbates) act as long-range electron scatterers
\cite{ando06,cheianov07,nomura07,hwang07,khveshchenko07}. It was shown
by several authors that Coulomb scattering (even when screened) leads
to a linear increase of the conductivity with electron density,
matching quite well the majority of the experimental data for
field-effect graphene-based devices. However, there are also a few
other scattering mechanisms that can also produce a linear dependence
of the conductivity on carrier density.

There is one additional type of disorder in graphene that, albeit
static, does not quite follow equation (\ref{eq:Hdis}). It is induced
by neutral adsorbates, such as atomic hydrogen, which tends to bind
covalently to carbon atoms in the graphene sheet and locally distort
both the lattice and electronic structure
\cite{boukhvalov08,wehling09}. The result is a mid-gap or resonant
state near the Dirac point that acts as a short-range scatterer,
similarly to (but not exactly like) vacancies. This type of disorder
also induces a dominant linear dependence of the conductivity on
carrier density, modulated by weak logarithmic term
\cite{stauber07,katsnelson07,titov,wehling10}. Scattering by ripples
in the graphene sheet can also, in principle, lead to a linear carrier
density dependence \cite{katsnelson08}, although this situation is
considered unlikely since it requires free-standing, equilibrium
fluctuations in the graphene sheet which are unlikely to dominate when
a substrate is present. Finally, a finite amount of wrinkles may also
induce the same linear dependence \cite{pereira10}.

There is some recent experimental evidence that resonant scatterers
may play a significant role in limiting mobility in graphene sheets
deposited on oxide substrates by mechanical exfoliation
\cite{ni10,katoch10}, but their microscopic nature is still
unclear. In fact, at present, the dominant mechanism of electron
scattering in graphene is still under debate for both suspended and
non-suspended graphene. In particular, for the latter, despite some
compelling evidence \cite{jang08}, the widespread view that Coulomb
impurities are the most important scattering mechanism limiting
mobility has been recently challenged \cite{ponomarenko09}.

Finally, we mention that disorder also plays a role in other
characteristics of graphene devices, such as $1/f$ noise
\cite{lin08,shao09,liu09}.

%%%%%%%%%%%%%%%%%%%%%%%%%%%%%%%%%%%%%%%%%%%%%%%%%%%%%%%%%%%%%%%%%%%

%%%%%%%%%%%%%%%%%%%%%%%%%%%%%%%%%%%%%%%%%%%%%%%%%%%%%%%%%%%%%%%%%%%
%% Neutrality point

% Graphene at the neutrality point
%

\section{Neutral graphene}
\label{sec:neutral}
%%%%%%%%%%%%%%%%%%%%%%%%%%%%%%%%%%%%%%%%%%%%%%%%%%%%%%%%%%%%%%%%%%%

Graphene at the neutrality point has some remarkable properties. As
mentioned earlier, the conductivity is finite even though, in the
clean limit, the density of states vanishes. Early transport
experiments found a conductivity minimum of the order of $e^2/h$,
which raised the possibility that this may actually be a universal
feature of graphene. It took some time for this issue to be settled
since several theories also pointed to some universality in the
conductivity of Dirac fermions at the neutrality point. For instance,
in the presence of certain types of chiral disorder such as random
vector potentials, it was well known that the conductivity is
essentially unaffected by disorder (as long as the disorder is not too
strong), taking the universal value of $4e^2/\pi h$
\cite{ludwig94,tsvelik95}. For other types of chiral disorder, the
conductivity at the neutrality point is also of order $e^2/h$
\cite{ostrovsky06} . In fact, more than two decades ago, Fradkin used
a coherent potential approximation to show that two-dimensional
zero-gap semiconductors have a universal Drude conductivity of order
$e^2/h$ at zero energy when intervalley scattering is neglected
\cite{fradkin86}. Some years later, Lee used a self-consistent Born
approximation (SCBA) to show that a similar result holds for fermions
at the nodal points of dirty d-wave superconductors \cite{lee93}. In
an early theoretical work, Shon and Ando used the SCBA to show that
for both short- and long-range impurity scattering, the conductivity
in graphene is equal to $2e^2/\pi h$ per spin degree of freedom
\cite{shon98} (see also \cite{ostrovsky06}). Coincidentally, this is
the same value of ballistic graphene when $W/L\gg 1$, as seen in
section \ref{sec:symmetry}.

As more experiments began to show that the conductivity minimum can in
fact vary widely from sample to sample \cite{tan07}, thus
contradicting the idea of universality, it became clear on the theory
side that the SCBA is inadequate to describe Dirac fermions near the
neutrality point. First, it does not correctly incorporate quantum
effects such as coherent scattering and localization. Second, it can
hardly be justified since at the Dirac point there is no clear
expansion parameter controlling the approximation (the carrier
wavelength diverges at the neutrality point for a uniform
system). Third, and perhaps more importantly, real systems are far
from being macroscopically homogeneous and uniform
\cite{martin08,zhang09,deshpande09,rossi08}. Strong fluctuations in
the background potential can occur at length scales comparable to the
mean free path $l$. This is not easily taken into account in the SCBA
or in any semiclassical formulation, although some recent attempts
have tried to circumvent this problem \cite{adam09a,adam07}.

The sample-dependent value of the conductivity minimum shows that the
nature and the strength of the disorder plays a crucial role in
determining the transport properties of graphene and a careful
theoretical analysis must be employed.

\begin{figure}[ht]
\centering
\includegraphics[width=10cm]{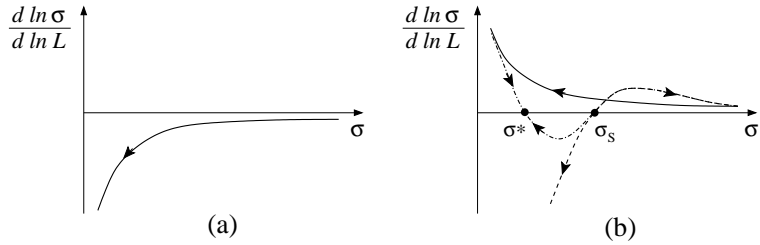}
\caption{One-parameter conductivity scaling functions proposed for
  graphene in the presence of (a) short-range and (b) long-range
  scalar disorder. In (b): the dashed curve represents the standard
  symplectic class with an unstable fixed point at $\sigma_S$, the
  dotted curve was conjectured in \cite{ostrovsky07a} and the solid
  curve corresponds to the result of \cite{ryu07,nomura07a} which is
  backed by numerical simulations
  \cite{bardarson07,sanjose07,lewenkopf08,tworzydlo08}. The arrows
  indicate the direction of the renormalization flow.}
\label{fig:beta}
\end{figure}

Let us first consider non-chiral, short-ranged potential
disorder. Scattering induced by localized defects, neutral impurities
or adsorbates can transfer enough momentum to carrier so that coupling
between valleys across the Brillouin zone (i.e. the ${\rm K}$ and
${\rm K}^\prime$ points) occurs. Several authors have considered this
regime theoretically within the renormalization group approach
\cite{khveshchenko06,aleiner06,altland06,ostrovsky06}. They have shown
that, as in the case of standard two-dimensional metals with no
spin-orbit coupling, coherent backscattering in graphene also leads to
carrier localization when the samples are sufficiently large, namely,
when the localization length $l_{\rm loc}$ is smaller than the sample
linear size $L$. Graphene in this case belongs to an orthogonal
symmetry class
\cite{ostrovsky07,khveshchenko06,aleiner06,altland06,mccann06}; in the
one-parameter scaling language, it has a beta function $\beta(\sigma)
= {\rm d}\ln\sigma/{\rm d}\ln L$ that is always negative, as shown in
figure \ref{fig:beta}(a). This result is supported by several
numerical simulations \cite{xiong07,lherbier08,amini09}. However, it
appears that, experimentally, $L$ is either too small or the dephasing
length $\ell_\phi$ is shorter than $l_{\rm loc}$ (we note that, in two
dimensions, the localization length can be exponentially larger than
the classical mean free path). Thus, the most common effect of
short-range disorder at the neutrality point is a reduction (but not a
complete suppression) of the conductivity. However, when short-range
disorder is strong, variable-range hoping conductivity is observed,
indicating localization. We will get back to short-range disorder when
we consider mesoscopic effects and doped graphene.

%%%%%%%%%%%%%%%%%%%%%%%%%%%%%%%%%
%%%%%%%%%%%%%%%%%%%%%%%%%%%%%%%%%

When the potential disorder is smooth at the atomic scale, intervalley
scattering (and therefore backscattering) is suppressed. As a result,
carriers can be described by a single-valley Dirac Hamiltonian with a
random scalar potential \cite{ludwig94,nersesyan95},
\begin{equation}
\label{eq:symplectic}
H = v_{\rm F}\, ({\bm p} \cdot \bsigma) + V({\bm r})\, \sigma_0.
\end{equation}
This Hamiltonian is symmetric under a pseudo time-reversal symmetry:
$H = \sigma_y\, H^T \sigma_y$ (notice that the real time-reversal
symmetry operation would connect ${\rm K}$ and ${\rm K}^\prime$ points
instead). This Hamiltonian belongs to a symplectic class
\cite{ostrovsky07,ludwig94,khveshchenko06,aleiner06,altland06,mccann06}. It
is well known that normal metals with spin-orbit coupling fall in this
class and show a metal-insulator transition \cite{hikami80},
therefore, it seems plausible that the same could occur for
graphene. In equation (\ref{eq:symplectic}), the pseudopsin (due to
the sublattice structure) plays a role similar to the real spin in
normal metal and is coupled to the orbital motion through the ${\bm p}
\cdot \bsigma$ term. From a theoretical viewpoint, until recently it
was not clear whether a transition also exists for graphene in this
class. Since it is difficult to fabricate samples with exclusively
long-range disorder, experiments did not provide any insight into this
question.

A field-theoretical treatment of this problem was proposed by two
groups \cite{ostrovsky07,ostrovsky07a,ryu07}. In particular, Ostrovky
{\it et al} have argued that the appropriate non-linear sigma model
includes a topological term which causes the appearance of a novel
stable fixed point at a finite conductivity $\sigma^\ast$ in addition
to the usual symplectic unstable fixed point $\sigma_S \approx 1.4
e^2/h$ \cite{ostrovsky07a,markos06}. The suppression of localization
was also obtained by Ryu {\it et al} \cite{ryu07} using a combination
of analytical and numerical calculations. These authors also derived a
topological term which is explicitly constrained to be either 0 or
$\pi$ by symmetry. In contrast to \cite{ostrovsky07a}, their treatment
of the global anomaly encoded in the topological term yields only one
stable metallic fixed point and no transition. Both results are
illustrated in figure \ref{fig:beta}(b) using a one-parameter scaling
beta function.

The derivation of the topological term in
\cite{ostrovsky07,ostrovsky07a} indicating the validity of the
single-parameter scaling of the conductivity, the existence of a
stable fixed point, and the consequent metal-insulator transition
requires a non-perturbative investigation. This motivated several
numerical studies that employed a variety of techniques
\cite{bardarson07,nomura07a,sanjose07,lewenkopf08,tworzydlo08}. Only a
strictly positive beta function was obtained and no evidence of a new
fixed point was found, thus corroborating the prediction of
Ref. \cite{ryu07}. This supports the idea that the main role of the
topological term is just to suppress the conventional symplectic
metal–insulator transition. In all studies, the average conductivity
followed a simple scaling law, increasing with the logarithm of the
system size such that $\beta(\sigma) =\alpha/\sigma$. While one could
in principle expect the constant $\alpha$ to be universal, the
situation is somewhat more complex, particularly for finite-size
systems (with the exception of \cite{nomura07a}), since simulations
are carried out at zero temperature and therefore in the fully
coherent limit. In addition, most simulations obtain the conductivity
by computing first the two-terminal conductance $G$ and can be
influenced by contacts and geometry.

%%%%%%%%%%%%%%%%%%%%%%%%%%%%%%%%%
\begin{figure}[ht]
\centering
\includegraphics[width=10cm]{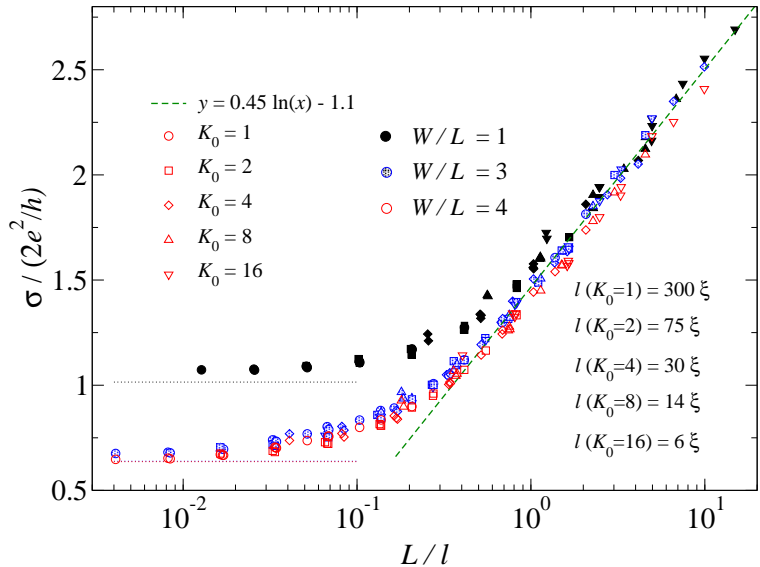}
\caption{Average conductivity of a graphene strip (armchair edges) as
  a function of rescaled length for three different aspect
  ratios. Several values of disorder strength ($K_0=1,2,4,8,16$) and
  correlation length ($\xi/a=2,4,8$) were used. The number of
  realizations range from 200 to 1000. The dashed line is a visual
  fitting to the large-$L$ portion of the numerical data. The dotted
  lines indicate the ballistic regime,
  equation (\ref{eq:pristineconductivity}).}
\label{fig:scaling}
\end{figure}
%%%%%%%%%%%%%%%%%%%%%%%%%%%%%%%%%

In our numerical simulations, we have found that the constant $\alpha$
deviates from the universal value of $e^2/\pi h$ expected for the
symplectic class. We use a single-band tight-binding model on a
honeycomb lattice with an added long-range Gaussian correlated
potential disorder \cite{ando06}, namely,
\begin{equation} 
\langle V({\bm r}) V({\bm r}^\prime)\rangle = K_0 \frac{(\hbar
  v_F)^2}{2\pi \xi^2} \exp \left(-\frac{|{\bm r} - {\bm
    r}^\prime|^2}{2\xi^2}\right),
\end{equation}
where $K_0$ parametrizes the disorder strength and $\xi$ is the
disorder correlation length. Figure \ref{fig:scaling} shows the
finite-size scaling of the average conductivity using the recursive
Green function method \cite{lewenkopf08}. In essence, this method
mimics a two-terminal transport measurement. Three different
aspect-ratios were considered as well as several values of $K_0$ and
$\xi/a$. The average conductivity obtained from the linear conductance
is plotted versus $L/l$, the ratio between system length and the
disorder mean free path $l$, which depends on $K_0$ and $\xi$. Two
clear regimes can be identified. For $L/l \ll 1$, the probability of
an electron being scattered by disorder as it traverses the sample is
very small. This corresponds to the ballistic regime, where scattering
occurs mainly at the sample edges and transport properties are
dominated by the sample geometry. In contrast, when $L/l\gg 1$, the
system becomes diffusive. Now, electrons are scattered multiple times
by disorder in their path across the sample. Scattering at the sample
edges is no longer dominant and transport depends weakly on the
geometry. Figure \ref{fig:scaling} clearly shows this crossover. In
the $L/l \rightarrow 0$ limit, the simulations approach the analytical
result (indicated by arrows) for clean graphene as given by equation
(\ref{eq:pristineconductivity}). For the diffusive regime, $L/l\gg 1$,
the conductivity is proportional to $\ln (L/l)$, in agreement with the
non-linear sigma model prediction \cite{ostrovsky06}. The mismatch
between the numerical prefactor for the logarithm and the value
characteristic of the symplectic class may be related to the finite
contact resistance.

These simulations suggest an explanation for the transport experiments
at the charge neutrality point. In the coherent diffusive regime, the
conductivity has significant sample-to-sample fluctuations and its
average shows a weak (logarithmic) dependence on the mean free
path. Typical diffusive experimental samples have $L/l \approx 1-10$
and $\sigma \approx 4e^2/h$, similarly to what is seen in figure
\ref{fig:scaling}.

An intuitive, non-rigorous, understanding of the lack of localization
in the presence of long-range disorder is possible by invoking the
so-called Klein tunneling
\cite{katsnelson06b,cheianov06,beenakker08}. In the absence of
intervalley scattering, as carriers move across charge puddles induced
by long-range potential fluctuations, chirality prevents
backscattering. When the Fermi energy cuts across a potential barrier,
carriers are converted from particle to hole (or
vice-verse). Chirality conservation then requires a forward moving
particle hitting the potential barrier at a normal angle to be
scattered as backward moving hole (interband tunneling). In this case
there is perfect {\it charge} transmission, as the real electron
involved in the process continues to move in the same direction. For
other incidence angles, a finite probability of particle
backscattering appears and it is actually a periodic function of the
angle. This effect has been recently observed experimentally
\cite{young09}. Klein tunneling through charge puddles (p- and n-doped
regions) is the basis of an attempt to explain the conductivity in
graphene near the neutrality point using a random resistor network
model \cite{cheianov07b}.

%%%%%%%%%%%%%%%%%%%%%%%%%%%%%%%%%%%%%%%%%%%%%%%%%%%%%%%%%%%%%%%%%%%

%%%%%%%%%%%%%%%%%%%%%%%%%%%%%%%%%%%%%%%%%%%%%%%%%%%%%%%%%%%%%%%%%%%
%% Mesoscopics

% Mesoscopics in graphene
%

\section{Mesoscopic effects}
\label{sec:mesoscopics}
%%%%%%%%%%%%%%%%%%%%%%%%%%%%%%%%%%%%%%%%%%%%%%%%%%%%%%%%%%%%%%%%%%%

Mesoscopic corrections to the conductivity have been studied for more
than 20 years in numerous systems, such as metals and two-dimensional
semiconductor heterostructures at low temperatures
\cite{datta95,imry97,ferry09,ihn09,nazarov09}. Weak localization and
universal conductance fluctuations are the most ubiquitous
manifestations of mesoscopic effects in transport. The weak
localization correction to the conductivity can be understood
semiclassically in terms of quantum interference between self-crossing
paths originated by disorder scattering. In the loops, an electron can
propagate in the clockwise and counter-clockwise directions and these
two trajectories interfere. The effect is maximal when the system is
time-reversal symmetric; for instance, in normal metals with
negligible spin-orbit coupling, it leads to constructive interference
and a decrease in the probability of carriers to move forward. This
interference effect gradually disappears as time-reversal symmetry is
broken, for instance, by an external magnetic field. In diffusive
systems, mesoscopic effects are also responsible for sample-to-sample
fluctuations in the conductance, $\delta G$, of order $e^2/h$,
irrespective of the magnitude of $G$ itself
\cite{reviewmesoscopics}. This phenomenon is known as universal
conductance fluctuations (UCF). The magnitude of quantum interference
effects strongly depends on $\ell_\phi/l$, the ratio between the
electron dephasing length and a sample specific length scale, such as
the electron elastic mean free path or the linear size of the sample
$L$. When $\ell_\phi \gg L$, this magnitude saturates to a constant
value; in convention metals, this value is solely determined by the
symmetry class and the sample geometry. Since, for any given sample,
$\ell_\phi$ decreases very rapidly with increasing temperature, low
temperatures are usually required to observe mesoscopic effects.

The effects of quantum interference in the electronic transport in a
honeycomb lattice had been theoretically addressed \cite{suzuura02}
even before graphene was first synthesized. Suzuura and Ando focused
their analysis on the weak localization peak, namely, the enhanced
resistivity at zero magnetic field. Using the standard disorder
diagrammatic approach for diffusive systems, which amounts to a
perturbative expansion in powers of $(k_{\rm F} l)^{-1}$, it was shown
that the weak localization correction depends strongly on the spatial
range of the disorder. (Notice that this expansion is only formally
justifiable away from the charge neutrality point, when $k_{\rm F}$ is
finite.) As mentioned in section \ref{sec:symmetry}, due to chirality,
backward scattering in graphene is suppressed for long-range
disorder. In this case, electrons in the vicinity of distinct valley
points do not couple and the system belongs to symplectic class,
leading to an anti-weak localization correction (negative
magnetoconductance), namely, a valley rather than a peak in the curve
of resistivity versus magnetic field. Short-range (non-magnetic)
disorder, in contrast, can break all symmetries except the
time-reversal one. In this situation, the system belongs to the
orthogonal class and there is weak localization (positive
magnetoconductance).

The theory was later extended to include trigonal warping effects
\cite{mccann06}. The diagrammatic expression for the weak localization
correction, $\Delta \sigma(B) = \sigma(B)-\sigma(0)$, as a function of
magnetic field reads
\begin{equation}
\label{eq:Dsigma}
\!\!\!\!\!\!\!\!\!\!\!\!\!\!\!\!\!\!\!\!\!\!\!\!\!\!\!\!
\Delta\sigma(B) = \frac{e^2}{\pi h} \left[ F
  \left( \frac{\tau_B^{-1}}{\tau_\phi^{-1}} \right) - F \left(
  \frac{\tau_B^{-1}}{\tau_\phi^{-1} + 2\tau_i^{-1}} \right) - 2 F
  \left( \frac{\tau_B^{-1}}{\tau_\phi^{-1} + \tau_i^{-1} +
    \tau_\ast^{-1}} \right) \right],
\end{equation}
where $F(z) = \ln z +\Psi(1/2 + 1/z)$ and $\Psi$ is the digamma
function. The magnetic field is cast in terms of $\tau_B^{-1} =
4eDB/\hbar$, where $D=v^2\tau_{\rm tr}/2$ denotes the diffusion
constant. The three other times scales correspond to: the dephasing
time $\tau_\phi$, which is related to the dephasing length as
$\ell_\phi = (D\tau_\phi)^{1/2}$; the intervalley scattering time
$\tau_{\rm i}$, which is due to sharp, atomic-like disorder features;
and the intra-valley scattering time $\tau_\ast$, which is related to
extended defects, dislocations, ripples, etc.. The derivation of this
expression assumes that the momentum relaxation time, $\tau_p$, is the
shortest of all time scales and is caused by charge impurities (and
not by atomically sharp defects), so that it does not affect
chirality.

The first experimental report on mesoscopic interference effects in
exfoliated graphene observed a strong suppression of the weak
localization peak, as compared to the typical value of $e^2/h$,
independently of doping even at very low temperatures
\cite{morozov06}. This behavior was attributed to the presence of
ripples in the graphene sheet: the effect of corrugations in the
electronic structure can be translated into an effective random gauge
field which destroys quantum interference in the same way as a random
magnetic field. Further experiments in different exfoliated graphene
samples \cite{tikhonenko08}, where temperature and carrier
concentration were systematically varied, observed much larger weak
localization peaks. Experiments in epitaxial graphene, on the other
hand, showed an extremely sharp anti-localization peak \cite{wu07}.

These observations are consistent with the theory of \cite{mccann06}
when the time scales appearing in equation (\ref{eq:Dsigma}) are
properly adjusted. In line with the qualitative discussion presented
above, a negative magnetoconductance correction or weak
anti-localization, $\Delta\sigma < 0$, corresponds to the case when
$\tau_i\rightarrow\infty$. Weak localization and positive
magnetoconductance, $\Delta\sigma>0$, occurs when $\tau_i\rightarrow
0$.

\begin{figure}[ht]
\centering
\includegraphics[width=5cm]{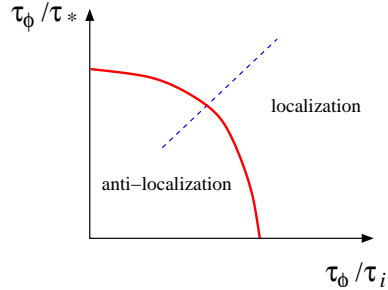}
\caption{Diagram illustrating the weak localization-weak
  anti-localization crossover. Adapted from \cite{tikhonenko09}.}
\label{fig:weak_loc}
\end{figure}

More interestingly, the theory predicts a region in the parameter
space where a transition from positive to negative magnetoconductance
corrections occurs. To see that, it is convenient to consider a weak
magnetic field, such that one can approximate the $F(z)$ function by
its leading order expansion, $F(z\rightarrow 0) \approx z^2/24$. Then,
\begin{equation}
\!\!\!\!\!\!\!\!\!\!\!\!\!\!\!\!\!\!\!\!\!
\Delta\sigma(B) \approx \frac{8\pi}{3} \frac{e^2}{h} \left(
\frac{B\ell_\phi^2}{\Phi_0} \right)^2 \left[ 1 -
  \frac{1}{(1+2\tau_\phi/\tau_i)^2} - \frac{2}{(1+\tau_\phi/\tau_i +
    \tau_\phi/\tau_\ast)^2} \right],
\end{equation}
where $\Phi_0 = h/e$ is the flux quantum. Based on this expression,
Tikhonenko {\it et al} \cite{tikhonenko09} proposed a diagram
reproduced in figure \ref{fig:weak_loc}. By following the dashed blue
line, one can cross from a localization to an anti-localization
region. Since the scattering times $\tau_i$ and $\tau_\ast$ are
essentially temperature independent but $\tau_\phi$ is not, one can
use temperature to move along this line. Another way to move in this
parameter space is to vary the carrier density: low density and high
resistivity is associated to short scattering times, while high
densities and low resistivity goes in the opposite direction. The
experimental data \cite{tikhonenko09} nicely support this analysis.

%%%%%%%%%%%%%%%%%%%%%%%%%%%%%%%%%
%%%%%%%%%%%%%%%%%%%%%%%%%%%%%%%%%

The picture that emerges from this analysis is that transport in
exfoliated graphene on SiO$_2$ substrates is influenced by both long-
and short-range scattering processes. For standard experimental
samples, the strength of short-range scattering processes is such that
$l_{\rm loc} \gg L$. Recent experiments \cite{chen09,moser10} show
that the conductivity is strongly suppressed when the concentration of
atomistically sharp defects is artificially increased, likely because
$l_{\rm loc} \ll L$.

\begin{figure}[ht]
\centering
\includegraphics[width=11cm]{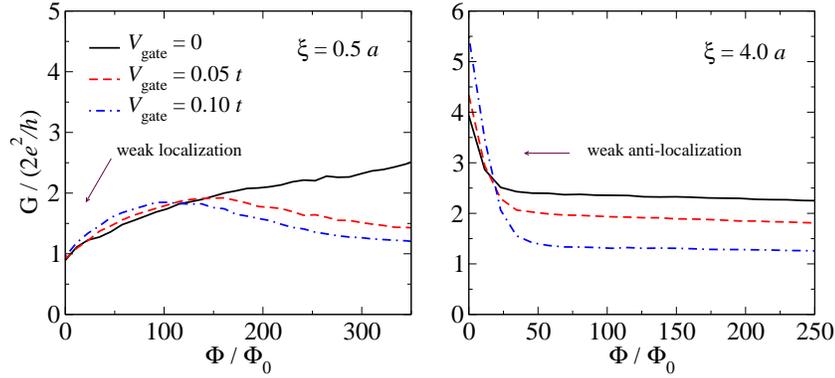}
\caption{Average magnetoconductance of an armchair strip ($W/L=3$) at
  ($V_{\rm gate} = 0$) and away from ($V_{\rm gate} > 0$) the
  neutrality point for short- ($\xi=0.5a$) and long-range ($\xi=4.0a$)
  disordered Gaussian potentials. $\Phi$ is the total magnetic flux
  perpendicular to the graphene sheet, $\Phi_0 = h/c$ is the flux
  quantum, and $t$ is the tight-binding nearest-neighbor hoping
  amplitude. A total of 500 realizations for each case have been used
  in computing the averages.}
\label{fig:finiteE_large}
\end{figure}

A quantitative study of mesoscopic quantum corrections to the
conductivity at the charge neutrality (Dirac) point lies outside the
validity range of the diagrammatic approach. The interplay between
localization length and the finite system length indicates that
numerical tools are necessary to gain further theoretical insight in
this case. Qualitatively, however, the main features are similar to
those obtained at finite doping since, particularly when charge
puddles are present, since in this case $k_{\rm F}$ is nonzero
locally. In figure \ref{fig:finiteE_large} we show the result of
numerical calculations of the magnetoconductance for a strip with
armchair edges, at the neutrality point and away, for both short- and
long-range disorder. The calculations were performed with a
tight-binding model and a Gaussian disordered potential, similarly to
those presented in figure \ref{fig:scaling}. Notice the transition
from weak localization to anti-localization as the range of the
potential is made larger than the lattice spacing.

%%%%%%%%%%%%%%%%%%%%%%%%%%%%%%%%%
%%%%%%%%%%%%%%%%%%%%%%%%%%%%%%%%%

Quantum interference among the multiple paths that carriers can take
as they move across the sample also cause conductance
fluctuations. Despite their random, aperiodic nature, these
fluctuations are reproducible when an external parameter is swept back
and forth. They are in essence quantum fingerprints of the microscopic
configuration of the disorder in the sample and are usually quite
evident when $L \ll \ell_\phi$. A proper way to quantify UCF is to
evaluate the autocorrelation function $C(\Delta X) = \langle \delta
G(X+\Delta X)\, \delta G(X)\rangle$, where $X$ represents an external
parameter such as a magnetic field and $\delta G$ is the deviation of
the conductance from its average value.

%Early simulations of UCF: Ando JPSJ 2004, Ando+Akimoto 2004,
%Akimoto+Ando 2004.???

Early transport measurements in graphene were able to identify
mesoscopic fluctuations
\cite{berger06,morozov06,heersche07,gorbachev07}, but only recently
more careful quantitative analyzes were carried out by several groups
\cite{staley08,horsell09,viera09,ojeda-aristizabal09,mason09}. On the
theory side, numerical simulations showed that long-range disorder in
graphene can produce anomalously large conductance fluctuations
\cite{rycerz07}, while for short-range disorder the magnitude of the
fluctuations is similar to that observed for normal diffusive metals.
This result was explained within the framework of the diagrammatic
perturbation theory by Kharitonov and Efetov \cite{kharitonov08}, who
highlighted the crucial role played by valley symmetry breaking in
determining the magnitude of the conductance variance ${\rm var}\, G =
C(0)$. For Dirac fermions, three possibilities arise: when all
contributions to scattering are relatively weak, ${\rm var}\, G$ is
four times larger than the standard value for normal diffusive
metals. When intravalley scattering is strong or triagonal warping is
present (high doping), ${\rm var}\, G$ is enhanced by just a factor of
two. Finally, when intervalley scattering is strong, ${\rm var}\, G$
takes the standard value. They also showed that when time-reversal
symmetry is broken by an external magnetic field, ${\rm var}\, G$ is
further reduced by a factor of two, as it is the case for conventional
metals. Similar results were also obtained in \cite{kechedzhi08},
where the role played by geometry (quasi-one-dimensional versus
two-dimensional) was also emphasized. However, as was the case for the
diagrammatic calculations of weak localization, both theories are only
applicable away from the neutrality point, as they rely on $k_{\rm
  F}l\gg 1$.

Experimentally, there is still some disagreement about the magnitude
and other characteristics of UCF in graphene. One group
\cite{staley08} reported that UCF are suppressed near the neutrality
point in both bilayer and trilayer graphene. The same effect was
observed in monolayer samples by the Illinois group \cite{mason09},
who related it to a marked reduction of the coherence length
$\ell_\phi$ near that point. However, another experiment on monolayer
and bilayer samples saw the opposite behavior
\cite{ojeda-aristizabal09} when mesoscopic fluctuations were collected
at a fixed value of an external magnetic field, while fluctuations
collected by varying the magnetic field were roughly independent of
the gate voltage. The authors in \cite{ojeda-aristizabal09} argued
that this discrepancy is due to the dominant presence of strong charge
inhomogeneities (puddles) in their samples: while varying an external
magnetic fields affect little the distribution of the puddles, gate
voltage variations are much more effective in scrambling the
conductance. There is also some discrepancy among the effect of
magnetic fields on the UCF amplitude: while \cite{horsell09} sees the
expected factor of two, in \cite{ojeda-aristizabal09} this only occurs
near the neutrality point, while in \cite{mason09} no reduction is
observed.

The temperature dependence of the UCF amplitude was measured by
several groups. In particular, \cite{horsell09} finds a clear inverse
power-law behavior at high doping, typical of conventional diffusive
metals, combined with a saturation at low temperatures. However, the
authors in \cite{viera09} observed an {\it exponential} suppression of
UCF with increasing temperature. While the former is backed by the
standard theory of dephasing in metallic systems \cite{altshuler}, the
latter has yet to be understood.

Finally, we mention a recent experiment where both weak localization
and UCF were observed when an {\it in-plane} magnetic field was
applied to graphene on SiO$_2$ \cite{lundberg09}. The appearance of
quantum interference in this case can be directly connected to
rippling in the graphene sheet.

%%%%%%%%%%%%%%%%%%%%%%%%%%%%%%%%%%%%%%%%%%%%%%%%%%%%%%%%%%%%%%%%%%%

%%%%%%%%%%%%%%%%%%%%%%%%%%%%%%%%%%%%%%%%%%%%%%%%%%%%%%%%%%%%%%%%%%%
%% Doped (finite density)

% Graphene at finite density of charge carriers
%

\section{Doped graphene}
\label{sec:doped}
%%%%%%%%%%%%%%%%%%%%%%%%%%%%%%%%%%%%%%%%%%%%%%%%%%%%%%%%%%%%%%%%%%%

Doped graphene shows remarkable high field-effect mobilities, even at
room temperatures \cite{novoselov04,morosov08}. This is a key property
that makes this material a potential candidate for future carbon-based
electronic devices. The dominant source of scattering limiting the
mobility $\mu$ in graphene is still under debate. As already
mentioned, at low temperatures, it is believed that charge impurities
\cite{ando06,nomura06,adam07,novikov07}, substrate effects, resonant
scattering \cite{stauber07,katsnelson07}, corrugations
\cite{katsnelson08}, and strain are the main sources of disorder.

With increasing doping, the conductivity of graphene depends only
weakly on temperature and grows almost linearly with carrier density,
in contrast with a $|n|^{1/2}$ dependence predicted for clean
graphene, equation (\ref{eq:ballistic}). These features were first
explained by considering charge impurity disorder, likely due to
charge trapped in the substrate
\cite{ando06,nomura06,adam07,novikov07}. These theories also predict
that the mobility should be inversely proportional to the impurity
concentration $n_{\rm imp}$ (for uncorrelated scatterers) and to
strongly depend on the dielectric constant. So far, there is no
experimental consensus on these properties. While experiments using
potassium ions provide evidence that $\mu \propto 1/n_{\rm imp}$
\cite{chen08}, experiments with gaseous adsorbates have only showed a
weak dependence of $\mu$ with $n_{\rm imp}$ \cite{schedin07}. There is
also an experimental controversy regarding the effect of the
dielectric constant on the conductivity: The Manchester group measured
only a modest change in $\mu$ by immersing graphene in high-$\kappa$
environments, such as ethanol and water \cite{ponomarenko09}, while
the Maryland group observed a significant change in $\mu$ by adding
just a few monolayers of ice to a graphene sheet
\cite{jang08}. Recently, experiments with hydrogenated graphene
revealed a strong short-range scattering component to the mobility,
apparently more dominant than charge impurities in some samples
\cite{ni10,katoch10}.

Let us now discuss the main theoretical approaches to address the
mobility in graphene. For finite doping such that $k_{F}l \gg 1$, it
is possible to calculate the conductivity within much a simpler
theoretical framework than that used at zero doping, where one deals
with a non-perturbative problem (see section \ref{sec:neutral}). In
addition, as one moves away from the charge neutrality point,
mesoscopic corrections to the conductivity lose their importance,
since $\delta \sigma/\sigma \ll 1$.

Assuming the system to be homogeneous and diffusive, $L\gg l$, the
simplest approach to describe the linear transport properties of
graphene is given by the Boltzmann theory
\cite{castroneto09,peres10,dassarma10}, which is widely used as a
guide for interpreting experimental results in doped graphene. This
semiclassical approach gives the conductivity in terms of the
transport scattering time $\tau_{\rm tr}$, namely
\begin{equation}
\sigma = \frac{e^2}{2}\int dE \left(-\frac{\partial f}{\partial
  E}\right)\nu(E)\, v_{\rm F}^2\, \tau_{\rm tr}(E),
\end{equation}
where $\nu(E)$ is the density of states, $f(E)$ is the Fermi-Dirac
distribution function, and $\tau_{\rm tr}$ is usually calculated using
Fermi's golden rule, namely,
\begin{equation}
\label{eq:momentum_relaxation}
\frac{\hbar}{\tau_{\rm tr}({\bm k})} = \frac{n_{\rm imp}}{4\pi}\int d
     {\bm k}^\prime |\widetilde{V}(q)|^2\left(1 - \hat{\bm k}\cdot
     {\hat{\bm k}}^\prime\right) F_{\hat{\bm k}{\hat{\bm k}}^\prime}
     \delta\big(E({\bm k}) - E({\bm k}^\prime) \big),
\end{equation}
where $q=|{\bm k} - {\bm k}^\prime| = 2 k \sin (\theta/2)$ is the
momentum transferred, $\widetilde{V}(q)$ is the Fourier transform of
the scattering impurity potential, and $\hat{\bm k}\cdot \hat{\bm
  k}^\prime = \cos\theta$. Chirality effects are captured by
$F_{\hat{\bm k}\hat{\bm k}^\prime}$, which is $F_{\hat{\bm k}\hat{\bm
    k}^\prime}=(1+\cos\theta)/2$ for graphene.

By modeling short-range disorder with zero-range delta-like scatters
$V({\bm r}) = \sum_i u_0 \delta({\bm r}-{\bm r}_i)$ \cite{shon98}, one
readily obtains that $\tau^{-1} \propto \nu(\epsilon)$. As a result,
the conductivity does not depend on the chemical potential $E_{\rm F}$
or on the carrier density $n$, which is obviously incompatible with
the experimental results. This observation is frequently used to
dismiss the importance of short-range scattering in graphene. Thus,
one has to bear in mind the limited applicability of this model. At
high doping and considering any kind of realistic of disorder with a
spatial range $\xi_{\rm range}$, such that $k_{\rm F} \xi_{\rm
  range}\neq 0$, modeling disorder by zero-range scatterers is hardly
justifiable. This model is also not very helpful for small doping
either, since the Boltzmann semiclassical approach breaks down for
small $k_{\rm F} l$. Even more importantly, it was realized very early
\cite{peres06} that short- (and nonzero-) range disorder leads to
resonant scattering. In this situation, the Fermi golden rule,
equation (\ref{eq:momentum_relaxation}), breaks down and $\tau_{\rm
  tr}$ should be calculated instead as follows \cite{ziman72}: (i) use
resonant scattering to calculate the transport cross section
$\sigma_T$, as it has been done, for instance, in \cite{peres10} for a
model where randomly placed hard discs of radius $R$ are the disorder
source. (ii) use $1/\tau_{\bm k} = n_{\rm imp} v_{\rm F} \sigma_T(kR)$
\cite{ziman72} instead of equation (\ref{eq:momentum_relaxation}). By
proceeding this way, it was shown
\cite{peres10,stauber07,katsnelson07} that resonant scatterers give
raise to mid-gap states and lead to a conductivity quasi-linear in
$n$, depending on the radius $R$ of the scattering discs.

The Boltzmann transport theory is customarily used to address the
effects of charge impurities in graphene
\cite{nomura06,ando06,hwang07}. In this case, the disorder scattering
potential $\widetilde{V}(q)$ becomes
$\widetilde{V}_C(q)/\varepsilon(q)$, where $\widetilde{V}_C(q) = 2\pi
e^2/(\kappa q)$ is the Fourier transform of the Coulomb potential in
an effective background dielectric constant $\kappa$, whereas
$\varepsilon(q)$ is the graphene dielectric function. Within this
disorder model, the influence of the substrate on the electronic
transport can be directly accessed by properly modifying
$\kappa$. This is the theoretical basis for the experimental work in
\cite{jang08,ponomarenko09}. The dielectric function
\begin{equation}
\label{eq:dielectric}    
\varepsilon(q, T) = 1 + v_C(q)\Pi(q, T)
\end{equation}
with $\Pi(q, T)$ denoting the irreducible polarizability function
given by the standard bare bubble diagram
\cite{ando06,wunsch06,hwang07}. The non-analytical behavior of
$\Pi(q)$ at $q=2k_F$ leads to Friedel-like oscillations in the
screened potential in real space, namely, $\phi(r) \sim
\cos(2k_Fr)/r^3$ \cite{cheianov06,wunsch06}. Both $\Pi(q, T)$ and the
chemical potential $\mu(T)$ temperature dependencies have a strong
influence on the conductivity. For low temperatures, such that
$T/T_{\rm F} \ll 1$, the conductivity depends very weakly on $T$,
whereas for $T/T_{\rm F}\gg 1$ one finds \cite{hwang09} that $\sigma
\propto (T/T_{\rm F})^2$.

The Boltzmann theory has been also used to calculate the conductivity
of graphene in the presence of static ripples \cite{katsnelson07} and
phonons \cite{stauber07,fratini08,mariani08}, with limited success in
explaining the experiments, in particular the temperature dependence
of $\sigma$. Let us also mention that the standard Boltzmann approach
was used to calculate the graphene conductivity for the correlated
Gaussian disorder model discussed in section \ref{sec:neutral},
yielding \cite{adam09}
\begin{equation} 
\label{eq:BoltzmannGaussian}
\sigma(n)= \frac{2\sqrt{\pi}e^2}{K_0 h} \left[ (2\pi |n| \xi^2)^{3/2}
  + O (|n|\xi^2)^{1/2}\right].
\end{equation}
Since $K_0 \propto n_{\rm imp}$ \cite{rycerz07}, the conductivity
increases linearly with the inverse disorder concentration $n_{\rm
  imp}^{-1}$. For high carrier concentrations, such that $n\xi^2 \gg
1$, $\sigma$ is proportional to $n^{3/2}$.

The next standard approximation level beyond the Boltzmann transport
theory is the self-consistent Born approximation, as mentioned in
section \ref{sec:neutral}. The starting point for the latter is the
Kubo formula, which gives the conductivity in terms of Green's
functions. While the standard Boltzmann approximation treats the
scattering within the Fermi Golden rule, the SCBA prescribes an
efficient way to encode the main scattering processes of an electron
in a disordered potential into the Green's function self-energy (see
for instance \cite{flensberg06}).

Ostrovsky {\it et al} \cite{ostrovsky06} presented a thorough study of
the conductivity due to generic (long- and short-range) Gaussian
disorder in monolayer graphene at finite bias using the
self-consistent Born approximation and a renormalization group (RG)
analysis \cite{aleiner06}. While for non-resonant (or weak) impurities
the SCBA and the RG analysis give similar results, for resonant
scattering the RG conductivity renders
\begin{equation}
\sigma(n)= \frac{e^2}{4\pi^2}\frac{|n|}{n_{\rm imp}}
\ln^2\frac{\Delta^2}{v_F^2 |n|}
\end{equation}
where $\Delta$ is a momentum cutoff. This result is very similar to
the one obtained for the hard disk model \cite{peres10}. Note that the
RG analysis for Gaussian disorder, including resonance scattering, is
significantly different than the one obtained using Fermi's golden
rule, equation (\ref{eq:BoltzmannGaussian}) \cite{adam09}.

The RG analysis \cite{aleiner06,ostrovsky06} has been also recently
employed to study the conductivity in the presence of randomly placed
adsorbates \cite{robinson08}. For resonant states far from the Fermi
energy, it predicts quite distinct and asymmetric curves for
$\sigma(n)$. For instance, $\sigma(n)$ can shown an almost
insulating-like behavior for p-type graphene and be metallic-like for
n-type graphene. This is consistent with some, but not all,
experiments with adsorbates.

Before concluding this section, it is worth mentioning that although
typical, good quality, micrometer-size graphene samples show $L/l>1$,
many do not belong to the regime $L/l\gg 1$. Hence, deviations from
the diffusive theory presented above, such as in \cite{tan07}, should
be expected. Numerical simulations provide insight into the
ballistic-diffusive crossover regime at finite doping
\cite{lewenkopf08}: with increasing disorder strength, the
conductivity dependence on the carrier concentration moves from a
sublinear dependence, resembling that of equation
(\ref{eq:ballistic}), to a superlinear dependence.

%%%%%%%%%%%%%%%%%%%%%%%%%%%%%%%%%%%%%%%%%%%%%%%%%%%%%%%%%%%%%%%%%%%

%%%%%%%%%%%%%%%%%%%%%%%%%%%%%%%%%%%%%%%%%%%%%%%%%%%%%%%%%%%%%%%%%%%
%% Conclusions

% Conclusions
%

\section{Conclusions and outlook}
\label{sec:conclusions}
%%%%%%%%%%%%%%%%%%%%%%%%%%%%%%%%%%%%%%%%%%%%%%%%%%%%%%%%%%%%%%%%%%%

In this review we attempted to highlight the most important
developments related to the role of disorder in the electronic
transport properties of graphene. We emphasized the role played by
symmetry breaking and discussed the various types of disorder commonly
found in exfoliated graphene sheets. Although some aspects of the
interplay between disorder and quantum coherence still require
theoretical and experimental investigation, much about strong and weak
localization and mesoscopic fluctuations of conductance in graphene is
already well understood. However, there is still no consensus about
which scattering mechanism plays the dominant role in limiting
mobility in both suspended and non-suspended graphene field-effect
devices. The widespread view that charge traps in the substrate are
the most effective scatterers of carriers in high-mobility graphene on
oxide substrates has been challenged by recent experiments.

Several important topics were not covered in this review. For
instance, disorder (intrinsic or extrinsic) plays an important role in
inducing spin-orbit coupling in graphene and, consequently, in
reducing spin relaxation times
\cite{huerta08,neto09a,popinciuc09,pi10}. Thus, investigating how
scattering mechanisms of various forms affect spin relaxation is
critical for advancing graphene as a spintronics material. Another
situation where disorder is critical is in nanoribbons
\cite{mucciolo09,evaldsson09,han10}. In general, it is fair to say
that there are still many open problems to explore in electronic
transport in disordered graphene systems.

% 
%- gap opening lattice distortions (strain)
%
%- disorder in nanoribbons (edges, coulomb block)
%
%- bilayers
%
%- ac conductivity
%
%- disorder in QHE
%
%- above all: numerical simulations can play a very important role in
%transport in graphene since large scale realistic modeling with
%atomistic, microscopic details is possible.

%%%%%%%%%%%%%%%%%%%%%%%%%%%%%%%%%%%%%%%%%%%%%%%%%%%%%%%%%%%%%%%%%%%

%%%%%%%%%%%%%%%%%%%%%%%%%%%%%%%%%%%%%%%%%%%%%%%%%%%%%%%%%%%%%%%%%%%
%% Acknowledgements

\ack

We would like to thank A H Castro Neto, M Ishigami, C Mudry, and N M R
Peres for useful discussions and V M Pereira for carefully reading the
manuscript. This work was supported in part by CNPq.

%%%%%%%%%%%%%%%%%%%%%%%%%%%%%%%%%%%%%%%%%%%%%%%%%%%%%%%%%%%%%%%%%%%
%% References

% References
%
%

%%%%%%%%%%%%%%%%%%%%%%%%%%%%%%%%%%%%%%%%%%%%%%%%%%%%%%%%%%%%%%%%%%%
\end{document}